\documentclass[fleqn,usenatbib]{mnras}

\usepackage{newtxtext,newtxmath}

\usepackage[T1]{fontenc}
\usepackage{ae,aecompl}

\usepackage{graphicx}	
\usepackage{amsmath}	
\usepackage{soul}

\newcommand{\bz}{$\langle B_z \rangle$}
\newcommand{\nz}{$\langle N_z \rangle$}
\newcommand{\dash}{NGC\,1624--2}

\title[UV variability and magnetic analysis of NGC 1624-2]{New observations of NGC 1624-2 reveal a complex magnetospheric structure and underlying surface magnetic geometry\thanks{Based on observations made at the Canada-France-Hawaii telescope (CFHT).}}

\author[A. David-Uraz et al.]{A. David-Uraz,$^{1,2,3}$\thanks{E-mail: alexandre.daviduraz@howard.edu}
V. Petit,$^{1,4}$
M.~E. Shultz,$^{1}$
A.~W. Fullerton,$^{5}$
C. Erba,$^{1}$\newauthor
Z. Keszthelyi,$^{6}$
S. Seadrow,$^{1}$
and G.~A. Wade$^{7}$
\\
$^{1}$Department of Physics and Astronomy, University of Delaware, Newark, DE 19716, USA\\
$^{2}$Department of Physics and Astronomy, Howard University, Washington, DC 20059, USA\\
$^{3}$Center for Research and Exploration in Space Science and Technology, and X-ray Astrophysics Laboratory, NASA/GSFC, Greenbelt, MD 20771, USA\\
$^{4}$ Bartol Research Institute, University of Delaware, Newark, DE 19716, USA\\
$^{5}$ Space Telescope Science Institute, Baltimore, MD 21218, USA\\
$^{6}$ Anton Pannekoek Institute for Astronomy, University of Amsterdam, Science Park 904, 1098 XH, Amsterdam, The Netherlands\\
$^{7}$ Department of Physics and Space Science, Royal Military College of Canada, PO Box 17000, Stn Forces, Kingston,
Ontario K7K 7B4, Canada
}

\date{Accepted 2020 December 1. Received 2020 December 1; in original form 2020 October 14}

\pubyear{2020}

\begin{document}
\label{firstpage}
\pagerange{\pageref{firstpage}--\pageref{lastpage}}
\maketitle

\begin{abstract}
NGC 1624-2 is the most strongly magnetized O-type star known. Previous spectroscopic observations of this object in the ultraviolet provided evidence that it hosts a large and dense circumstellar magnetosphere. Follow-up observations obtained with the \textit{Hubble Space Telescope} not only confirm that previous inference, but also suggest that NGC 1624-2's magnetosphere has a complex structure. Furthermore, an expanded spectropolarimetric time series shows a potential departure from a dipolar magnetic field geometry, which could mean that the strongest field detected at the surface of an O-type star is also topologically complex. This result raises important questions regarding the origin and evolution of magnetic fields in massive stars.
\end{abstract}




\section{Introduction}\label{sec:intro}

As sources of ionizing flux, momentum and chemical enrichment, O stars represent formidable galactic 
engines, sculpting their immediate environment and participating in the overall evolution of their galactic hosts \citep[e.g.][]{2003IAUS..212..585C}
. Of particular interest is the subset ($\sim$7 per cent; \citealt{2017MNRAS.465.2432G}) of O stars that exhibit a detectable surface magnetic field, thought to potentially be the progenitors of exotic objects and transients such as heavy stellar-mass black holes \citep{2017MNRAS.466.1052P, 2020ApJ...900...98G} and pair instability supernovae \citep{2017A&A...599L...5G} even at solar metallicity, as well as magnetars \citep{2008MNRAS.389L..66F}
.

There are many limitations to our understanding of these stars and of the role that they play in galactic ecology. The most obvious one concerns their rarity: only 11 such objects are known to exist so far \citep{2013MNRAS.429..398P, 2016A&A...592A..84F}, making an in-depth characterization of their magnetic properties as a population 
difficult due to small number statistics. Furthermore, these stars have undergone ``magnetic braking" \citep{1967ApJ...148..217W, 2009MNRAS.392.1022U} and hence are typically slow rotators, which means their detailed study requires long monitoring programs 
in some cases. Indeed, the O-type star with the longest known rotation period is the magnetic O4-8f?p star HD 108, with a period of $\sim$55 yr \citep{2010A&A...520A..59N, 2017MNRAS.468.3985S}. Even for those few stars 
which do have shorter rotation periods, other complications (such as binarity) 
muddle the interpretation of their observations 
and properties (e.g. HD 47129, also called ``Plaskett's star"; \citealt{2013MNRAS.428.1686G}).

That said, as far as we 
know from the results of large-scale spectropolarimetric surveys (e.g. \citealt{2015IAUS..307..342M, 2016MNRAS.456....2W}), magnetic O stars harbour essentially dipolar, globally-organized fields on their surfaces, with strengths 
on the order of a few kG -- with the notable exceptions of $\zeta$ Ori A 
and NGC 1624-2 
on the lower ($\sim$140 G; \citealt{2015A&A...582A.110B}) and higher ($\sim$20 kG; \citealt{2012MNRAS.425.1278W}) end of this range, respectively. Recent studies suggest that O stars might not follow the long-held assumption of magnetic flux conservation as they evolve (\citealt{2007A&A...470..685L,2008A&A...481..465L,2016A&A...592A..84F,2019MNRAS.490..274S}; though large surveys may not as yet conclusively address this issue; \citealt{2019MNRAS.489.5669P}), although the details of the dissipation mechanism and rate remain unknown.

One key characteristic of magnetic O stars is the interplay between their dense, radiatively-driven winds (e.g. \citealt{1975ApJ...195..157C}) and their magnetic fields, which channel and confine their outflows to form circumstellar \textit{magnetospheres} \citep{1978ApJ...224L...5L, 1990ApJ...365..665S, 1997ApJ...485L..29B}. Early magnetohydrodynamic (MHD) simulations provided valuable insights into the structure of magnetospheres, as well as their role in spinning down their host stars and inhibiting their gradual loss of mass throughout their main-sequence lifetimes \citep{2002ApJ...576..413U,2008MNRAS.385...97U,2009MNRAS.392.1022U}. Incorporation of these 
effects 
into dedicated evolutionary models 
has profound consequences on the end points of the evolution of magnetic massive stars 
\citep[e.g.][]{2011A&A...525L..11M,2019MNRAS.485.5843K,2020MNRAS.493..518K}.

Given that most magnetic O stars have very 
low rotation rates (as a consequence of the braking mentioned above; \citealt{2013MNRAS.429..398P}), the confined wind material within their magnetospheres is not centrifugally supported and therefore the analytic prescription of \citet{2016MNRAS.462.3830O} -- or \textit{Analytic Dynamical Magnetosphere} (ADM) model -- presents a practical time-averaged view of the structure of the magnetosphere that agrees well with earlier MHD investigations. Such a prescription is ideally suited to the interpretation of various multiwavelength magnetospheric diagnostics (such as H$\alpha$; \citealt{2016MNRAS.462.3830O}; X-ray emission; \citealt{2014ApJS..215...10N}; optical broadband photometry; \citealt{2020MNRAS.492.1199M}; as well as ultraviolet resonant line profiles; \citealt{2019arXiv191208748E}), without the computational cost associated with full-scale MHD simulations. Indeed, while 
none of these diagnostics 
individually contains sufficient information to construct a fully model-independent view of the magnetosphere (such as e.g. a 3-dimensional tomographic Doppler reconstruction of the magnetospheric structure), 
qualitative inferences can be made in light of simple analytic tools such as the ADM model.

In particular, wind-sensitive ultraviolet (UV) resonance lines 
have historically been coupled very successfully with spherically symmetric models to provide important insights into the mass-loss properties of the general population of O stars (e.g. \citealt{2003ApJ...595.1182B}). Their ability to probe the density and velocity structure of the extended atmospheres of these stars makes them particularly useful in that regard, and that usefulness carries over to magnetic stars as well, although the assumption of spherical symmetry breaks down and dedicated models must be used to interpret their variations. For instance, assuming a globally-organized dipolar field, it is fairly 
simple to distinguish between different viewing angles
. The general phenomenology of the UV line profile variations (e.g. \citealt{2013MNRAS.431.2253M}) can be framed in terms of the rotational phases associated with the ``high" ($\phi = 0.0$) and ``low" ($\phi = 0.5$) states, which 
are defined respectively by the maximum and minimum of the 
variation exhibited by the equivalent width of H$\alpha$.
In the context of the oblique rotator model \citep{1950MNRAS.110..395S}, variations in the equivalent width of H$\alpha$ are attributed to changes in the projected
surface area of the region of enhanced magnetospheric emission that forms 
in the magnetic 
equatorial plane.
Consequently, the ``high" state corresponds to times when the magnetic pole is most closely aligned to the observer's line of sight, 
while the ``low" state occurs when the magnetic equator is closest to being viewed edge on. Within that framework, ``strong" UV resonance lines (i.e. with large oscillator strengths, and thus nearly optically thick when formed in dense O-star winds) show enhanced high-velocity absorption at high state, since we are looking 
through wind material flowing nearly radially along open magnetic field lines, while enhanced absorption can be observed near line center in both strong and weaker lines 
in the low state (e.g., \citealt{2015MNRAS.452.2641N}).

The archetype of an O star magnetosphere can be found around the Of?p\footnote{This spectral class was defined by \citet{1972AJ.....77..312W} and found to be related to magnetism by \citet{2017MNRAS.465.2432G}.}  
star NGC 1624-2 (O7f?p, \citealt{2010ApJ...711L.143W}). Discovered to host the strongest known magnetic field at the surface of an O-type star by almost an order of magnitude ($\sim$20 kG, assuming a dipolar geometry; \citealt{2012MNRAS.425.1278W}), it is also the only star of that peculiar spectral type to exhibit clearly separated Zeeman components in its integrated Stokes $I$ 
line profiles. This allows for a complementary measure of the disc-integrated field modulus (in addition to the usual disc-integrated longitudinal field measurements enabled by circular polarization and typically used to diagnose magnetic fields in massive stars). Preliminary attempts (which we revisit in this paper with additional monitoring) to reconcile both field modulus and longitudinal field measurements have so far failed to do so using 
a solely dipolar field
, hinting at a 
more complex magnetic geometry (\citealt{2012MNRAS.425.1278W, 2016MSc.........16M, 2017IAUS..329..394D}).

Previous X-ray and UV investigations have found NGC 1624-2 to be surrounded by a giant magnetosphere, as expected from its large Alfv\'{e}n radius ($R_A/R_* \sim 11.4$, \citealt{2012MNRAS.425.1278W}), the largest of any O-type star. \citet{2015MNRAS.453.3288P} showed that it must be large and dense, as it absorbs up to 95 per cent of the intrinsic circumstellar X-ray emission, a result 
further supported by \citet{2019MNRAS.483.2814D}, who found it to cause very large line profile variations in the UV.

In this paper, we present new UV spectra taken with the \textit{Hubble Space Telescope} (\textit{HST}) as well as spectropolarimetric observations taken at the Canada-France-Hawaii Telescope (CFHT) and the T\'{e}lescope Bernard-Lyot (TBL)
. In Section~\ref{sec:obs}, we provide more details about the observations, while the 
UV and magnetic results are presented in Section~\ref{sec:uv}
. Finally, we 
interpret these results and offer our conclusions in Section~\ref{sec:concl}.

\section{Observations}\label{sec:obs}

\subsection{Ultraviolet spectroscopy}
\label{sec:hst} 


Previous observations of {\dash} were obtained by GO Program 13734 (PI: Petit; \citealt{2019MNRAS.483.2814D}) with the Cosmic Origins Spectrograph (COS) 
on board the Hubble Space Telescope (\textit{HST}), and
sampled the morphology of the 
key UV resonance lines at the phases of the 157.99 day period associated with the ``high" ($\phi \sim 0.0$) and ``low" ($\phi \sim 0.5$) states
.
General Observer Program 15066 (PI: David-Uraz) also used 
COS to obtain 4 additional ultraviolet spectra of {\dash}  near the two quadrature phases, in order to assess asymmetries in the
structure of the magnetosphere.
Table~{\ref{tab:obs}} summarizes the available UV spectroscopy of {\dash}.

Each COS observation obtained by GO Program 15066 consisted of a two-orbit visit.
For each visit, the first orbit began with a guide-star acquisition by the Fine Guidance Sensor (FGS) 
followed by a target acquisition using the imaging mode of COS with the near-UV channel, Mirror A, and the
bright-object aperture (BOA). 
After the execution of a small-angle maneuver to center {\dash} accurately in the primary science aperture
(PSA; $2.5$\arcsec \ diameter),  spectra with the G130M grating centered on 1222~{\AA} and 1291~{\AA} were obtained. 
During the second orbit, exposures with the G160M grating centered on wavelengths of 1577~{\AA} and 1623~{\AA} were acquired. 
Both channels of the far-ultraviolet cross-delay (XDL) line detector were used to make the observations in time-tag mode.

These four instrumental configurations provided useful wavelength coverage from about 1150 to 1800~{\AA} with a resolving power 
that increases linearly with wavelength from $\sim$16,000 to $\sim$21,000 over the respective ranges of both gratings. 
For each grating setting, the total exposure time indicated in Table~{\ref{tab:obs}} was divided into sub-exposures of equal duration
but slightly different positions (FP-POS) on the detector to mitigate the effects of fixed-pattern noise
in the combined ``x1dsum" spectrum.
Since the observations were made at COS lifetime position 4, only 2 FP-POS positions were available for the G130M/1291 configuration.  
For all other configurations, 4 FP-POS positions were used.

The spectra were uniformly processed with version 3.3.4 (2018-01-08) of the CALCOS calibration pipeline.
Calibration steps included: correcting the photon-event table for dead-time and positional
effects such as drifts in the detector electronics, geometric distortion, and the
Doppler shift of the observatory; binning the time-tag data and assigning wavelengths
to the bins on the basis of Pt-Ne spectra acquired simultaneously with spectra of
{\dash}; extraction and photometric calibration of 1-D spectra; and the ``shift and add"
combination of spectra taken at different FP-POS positions for each grating setting.



\subsection{Optical spectropolarimetry}

NGC 1624-2 was observed with 
ESPaDOnS (an Echelle SpectroPolarimetric Device for the Observation of Stars; \citealt{2006ASPC..358..362D}) at the CFHT, as part of observing programs 12AP14, 12BP13 and 13BC05 (PI: Wade), as well as program 15BC13 (PI: Petit). This instrument has a high 
{resolving power} ($R \sim 65000$) and covers a wavelength range of about 3600-10000 \AA. In total, 22 observations (including 5 that were previously analysed by \citealt{2012MNRAS.425.1278W}), each consisting of four subexposures, were obtained between 2012 and 2015. Data reduction was performed using the \textsc{Upena} pipeline \citep{2011tfa..confE..63M} 
based on the \textsc{Libre-ESpRIT} reduction package \citep{1997MNRAS.291..658D}. It combines the aforementioned subexposures (each corresponding to a different angle of the Fresnel rhombs) to yield integrated spectra (Stokes $I$ parameter), as well as circularly polarized spectra (Stokes $V$ parameter), which are sensitive to the Zeeman effect 
\citep[e.g.][]{2004ASSL..307.....L}, thus allowing us to detect and measure astrophysical magnetic fields
. 
It also produces two diagnostic nulls ($N$), 
which characterize the level of noise and allow us to identify potential spurious signals in the Stokes $V$ spectra
.

We also included an additional observation obtained with Narval (a twin instrument) using the TBL (program L121N02), which was included in the analysis of \citet{2012MNRAS.425.1278W}. These observations are detailed in Table~\ref{tab:Obs_mag}, together with mean longitudinal field measurements, while 
separate longitudinal field measurements obtained from single lines (see Section~\ref{sec:magfield}) appear in Table~\ref{tab:line_bz}. 

\section{Results}\label{sec:uv}
\subsection{UV analysis}

The new spectra were continuum normalized. 
In particular, we focused on the profiles of the same wind-sensitive lines 
that were studied by \citet{2019MNRAS.483.2814D}, namely the Si~\textsc{iv}$\mathrm{\lambda}\mathrm{\lambda}$1393/1402 and C~\textsc{iv}$\mathrm{\lambda}\mathrm{\lambda}$1548/50 doublets. These lines probe two important regimes of opacity, and in the context of a simple dipolar magnetic field, their variation can be understood in terms of the geometry (density and velocity) of the magnetospheric plasma. Indeed, building upon the picture presented in Section~\ref{sec:intro}, the general phenomenology can be understood as follows: the unconfined wind material along the magnetic poles escapes nearly radially at high velocity and is rarefied, while around the magnetic equator, the dense magnetically-confined material moves at a lower velocity, and mostly in the azimuthal direction (with respect to the magnetic axis). Therefore, a weaker line like the Si~\textsc{iv} doublet probes mostly high-density regions near the stellar surface, leading to low-to-moderate velocity emission at high state, when the dense material trapped near the equator is seen off-limb, and to strong absorption near line center (and \textit{stronger} absorption generally) when that same material occults the star at low state. Additionally, a stronger line such as the C~\textsc{iv} doublet is also sensitive to the rarefied near-terminal\footnote{\citet{2012MNRAS.425.1278W} report NGC 1624-2's terminal velocity to be 2875 km\,s$^{-1}$, based on a theoretical scaling \citep{1995ApJ...455..269L} with its escape velocity. This represents, in principle, the terminal velocity that the star would have, with its other properties (mass, radius, luminosity) unchanged, in the absence of a magnetic field. However, the interaction with the magnetic field at the base of the wind can modify its overall driving, as evidenced by the azimuthal dependence of the surface mass flux in MHD simulations of magnetic massive stars \citep{2004ApJ...600.1004O}. Moreover, these same simulations also show that the polar flow can reach velocities greater than the nominal terminal velocity, due to additional driving allowed by the faster-than-radial divergence of the flux tubes, leading to rarefied material and consequently to desaturation in the high-velocity absorption trough of strong wind lines. Therefore, while this reported value of the terminal velocity should only be used as an estimate, a more profound understanding of the kinematics of the wind can only be obtained by fitting the UV wind-sensitive line profiles of NGC 1624-2 with appropriate, non-spherically-symmetric models (e.g. \citealt{2019arXiv191208748E}).} wind streaming above the magnetic pole, which means that there should be additional high-velocity blueshifted absorption at high state. This picture corresponds to the general case described by \citet{2015MNRAS.452.2641N}, as well as what was observed by \citet{2019MNRAS.483.2814D}, who also pointed out that the Si~\textsc{iv} and C~\textsc{iv} line profiles in the spectra of NGC 1624-2 exhibit overall more emission than those of comparable non-magnetic stars (due to the overdense confined magnetic equatorial region, compared to a spherically-symmetric wind) and much greater variability than even those of other magnetic O-type stars.

\begin{figure*}
	\includegraphics[width=6.5in, trim=85 140 100 150, clip]{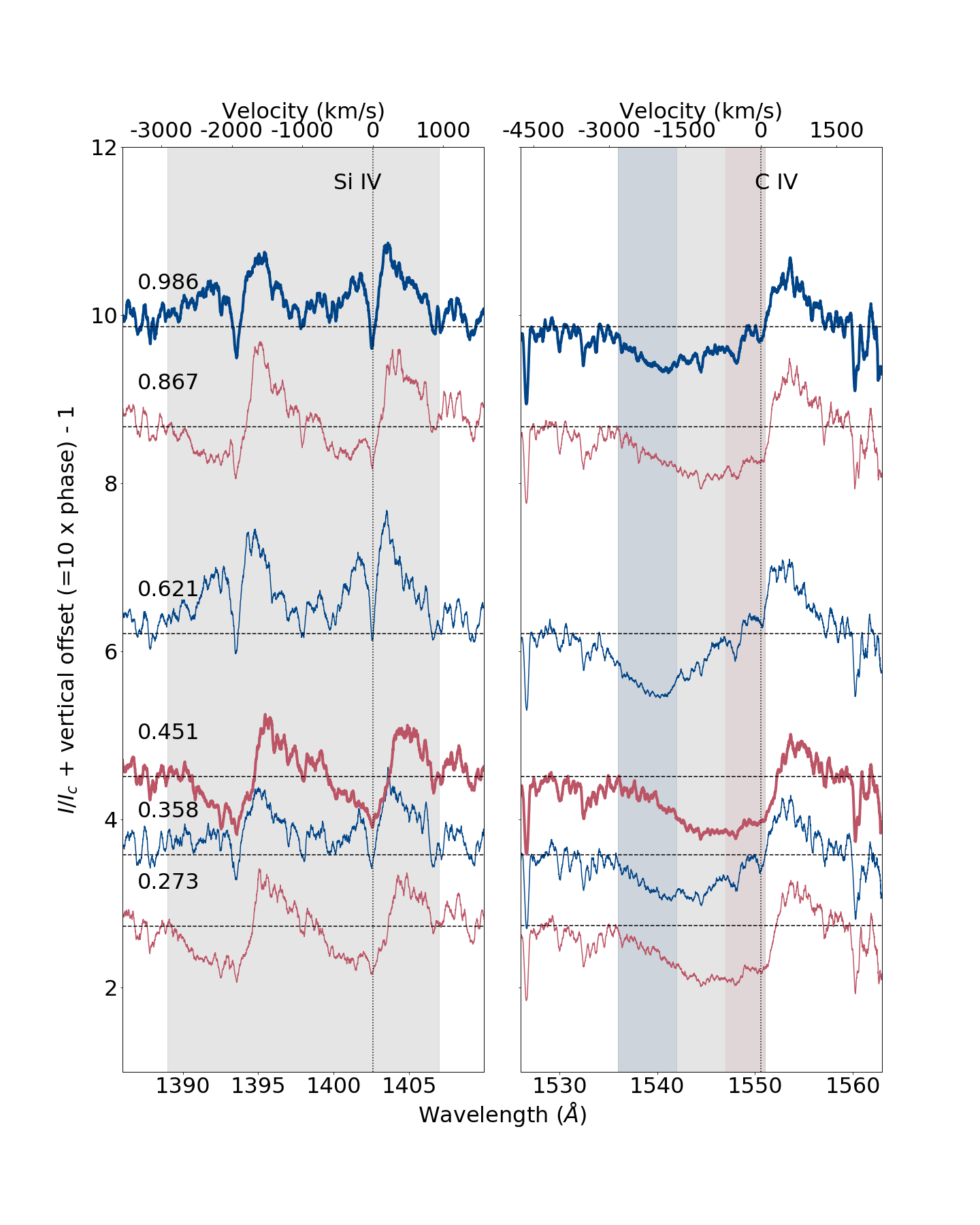}
    \caption{\textit{HST/COS} spectra of NGC 1624-2 obtained at 6 different rotational phases, as computed using the ephemerides published by \citet{2012MNRAS.425.1278W}. Each spectrum is vertically offset proportionally to its phase, which is included as a label. Magenta lines represent spectra that have ``low-state-like" behaviour, while blue lines represent ``high-state-like" spectra. The thicker lines correspond to spectra that were already presented by \citet{2019MNRAS.483.2814D}. These spectra were smoothed with a 21-pixel wide boxcar filter for display purposes. Vertical background bands correspond to ranges of integration for our equivalent width calculations, with the grey bands corresponding to the whole range of the Si~\textsc{iv} doublet and of the absorption trough of the C~\textsc{iv} doublet, and on the right, the overlaid magenta and blue bands correspond 
    to the low- and high-velocity portions of the aforementioned absorption trough
    , which respectively characterise low-state-like and high-state-like morphological features. The top axis shows the corresponding velocities for each panel, in the reference frame of the star, with respect to the line center of the red component of the doublet (indicated by a vertical dotted line).}
    \label{fig:UVspec}
\end{figure*}

A comparison of the newly-obtained profiles with 
those obtained previously by \citet{2019MNRAS.483.2814D} is presented in Fig.~\ref{fig:UVspec}. While one would normally expect to see -- assuming that the magnetosphere is formed by a global dipolar field -- a gradual transition (e.g. \citealt{2019arXiv191208748E}) between the 
high state 
and the 
low state 
profiles, and vice-versa, that is not 
what is observed.

Instead, 
every newly-observed line profile 
closely resembles 
one of the previous two observations. In particular, 
observations obtained at rotational phases $\sim$0.36 and $\sim$0.62 are very similar to the previous 
high-state 
observation obtained at phase $\sim$0.99, in that they present enhanced emission in the Si~\textsc{iv} doublet as well as enhanced absorption at high velocities (i.e. in the blue vertical band in Fig.~\ref{fig:UVspec}) on the blue side of the C~\textsc{iv} doublet. Similarly, 
observations obtained at phases $\sim$0.27 and $\sim$0.87 are very similar to the previous 
low-state 
observation obtained at phase $\sim$0.45, with stronger absorption close to line center in both doublets. In fact, the observation at phase 0.62 appears to exhibit even more extreme high-state line profile features than that at phase 0.99, with even stronger emission at low velocities in the Si~\textsc{iv} doublet and deeper absorption at high velocities on the blue side of the C~\textsc{iv} doublet.

In principle, one might have expected that either the high-state and low-state do not quite correspond to magnetic pole-on and equator-on views, respectively, or that the viewing angle crosses the magnetic equator more than once over the course of a rotational cycle, leading to double-wave variations of the line profile 
variability described above. That said, neither one of these scenarios can explain: (i) the fact that observations taken at \textit{three} separate rotational phases show similar line morphologies, and (ii) that the ``high-state-like" and ``low-state-like" morphologies alternate throughout a rotational cycle. This is also at odds with the behaviour of other magnetospheric diagnostics, 
such as H$\alpha$ emission \citep{2012MNRAS.425.1278W}
.


To further illustrate this point, we compute equivalent widths using the wavelength ranges shown in Fig.~\ref{fig:UVspec}: the full ranges of the Si~\textsc{iv} doublet and of the absorption trough of the C~\textsc{iv} doublet, as well as two separate ranges within the blue side of the C~\textsc{iv} doublet corresponding to high/low velocities. These ranges were chosen to best illustrate the interplay between emission and absorption effects, based on the phenomenology described above. 
The results of our measurements are shown in the top four panels of Fig.~\ref{bz_fwhm}. We can see that for a given integration range, all three equivalent width measurements taken at low-state-like phases (represented by downward-pointing red triangles) are consistent with each other within the error bars. Similarly, the high-state-like measurements (represented by upward-pointing blue triangles) are also relatively consistent with each other, with the exception of the point at phase 0.621, especially for the high-velocity absorption on the blue side of the C~\textsc{iv} doublet. They are also clearly distinct from the low-state-like measurements, except once again in the case of the high-velocity absorption in the C~\textsc{iv} doublet, for which the difference between both sets of measurements is 
not significant within the errors, though systematic. Furthermore, the overall behaviour corresponds to what can be seen from the line profiles themselves and to the general phenomenological picture presented above, although the phasing cannot be explained in the context of a dipolar magnetic field (as also evidenced by the difference between these equivalent width curves and those of other magnetic O-type stars, such as HD 191612, thought to have a dipolar magnetic field; \citealt{2013MNRAS.431.2253M}). The Si~\textsc{iv} doublet shows greatly enhanced emission at high-state-like phases, presumably due to dense confined material seen off the limb of the star. Meanwhile, the high-velocity absorption trough of the C~\textsc{iv} doublet shows deeper absorption when the star is viewed in such a way that fast-flowing wind material along open field lines intersects the line of sight (high-state-like phases), and conversely the low-velocity absorption near line center is enhanced when an accumulation of wind material confined near the surface occults the stellar disc (low-state-like phases). These combined effects lead to enhanced absorption at low-state-like phases in the full range of the absorption trough of the C~\textsc{iv} doublet. Additional UV diagnostics also show a similar dichotomy, such as the ``forest" of photospheric Fe~\textsc{iv} lines in the 1615-1630 \AA{} range, which show enhanced absorption at low-state-like phases, as found by \citet{2019MNRAS.483.2814D}.


In principle, it appears as though these observations might provide some support for the hypothesis of a complex field (i.e. departing from a simple dipole), as investigated in greater detail below. We also attempt to develop further inferences on the global structure of the magnetosphere in Section~\ref{sec:mag}.

\subsection{Magnetic analysis}\label{sec:magfield}

The optical spectra were continuum normalized and co-added (for spectra taken 
within a few days of each other, hence similar rotational phase), and then used to measure the 
disk-averaged line-of-sight magnetic field \bz~\citep[as defined by][]{mat1989} 
from a selection of spectral lines chosen to be strong, isolated and relatively uncontaminated by wind emission. The laboratory wavelengths and Land\'e factors (given in Table \ref{tab:line_bz}, along with the measurements from each line) were obtained from a line list downloaded from the Vienna Atomic Line Database \citep[VALD3;][]{piskunov1995, ryabchikova1997, kupka1999, kupka2000,2015PhyS...90e4005R} with an `extract stellar' request formulated with NGC\,1624-2's stellar parameters ($T_\textrm{eff}$ = 35 kK, $\log g$ = 4.0; \citealt{2012MNRAS.425.1278W}). The null field \nz~was also obtained from the diagnostic $N$ profile in the same fashion; \nz~is consistent with 0 in all measurements. Integration ranges were determined using the test described by \cite{neiner2012b}, i.e.\ progressively widening the integration range until \bz~converges to a constant value, and selecting the integration range yielding the minimum error bar.

\begin{figure}
\includegraphics[width=0.49\textwidth, trim =20 50 20 40, clip]{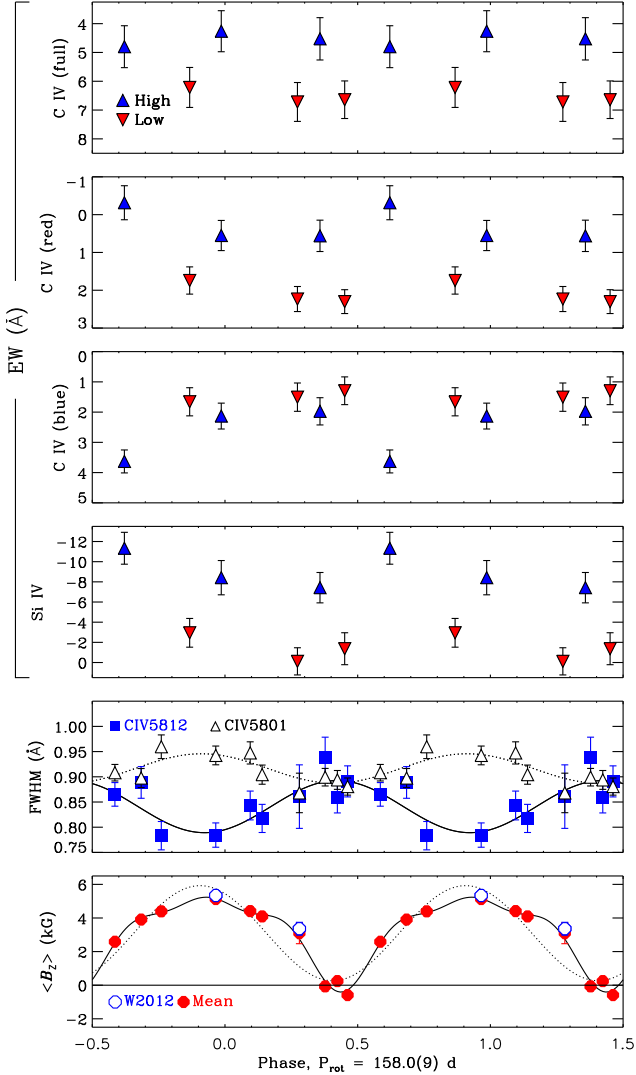}
\caption{{\em Top 4 panels}: Equivalent widths, folded with the rotation period, of (in order from the top) the blue side of the C~{\sc iv} profile, the low- and high-velocity portions of the blue side of the C~{\sc iv} doublet, and the Si~\textsc{iv} doublet, within the ranges defined in Fig.~\ref{fig:UVspec}. Observations at high- and low-state-like phases are indicated by upward-pointing blue and downward-pointing red triangles, respectively. We can see that all three low-state-like measurements agree within errors for each of the three wavelength ranges, while the high-state-like measurements have different values and are relatively consistent with each other (with the point at phase 0.621 exhibiting slightly more extreme behaviour). {\em Fifth panel}: FWHM of the C~{\sc iv} $\lambda$5801 and $\lambda$5812 lines, as a proxy to the magnetic modulus. Solid and dotted curves are harmonic fits. {\em Bottom panel}: Weighted mean \bz~measurements across all lines (including those yielded by the spectra used by \citet{2012MNRAS.425.1278W}, in blue, which we reanalyzed for uniformity -- our results are consistent, within errors, with theirs). Solid and dotted curves show third- and first-order harmonic fits, respectively.}
\label{bz_fwhm}
\end{figure}

The bottom panel of Fig.\ \ref{bz_fwhm} shows \bz~folded with the 157.99~d rotation period \citep{2012MNRAS.425.1278W}. As can be seen in Fig.\ \ref{bz_fwhm}, where we show the weighted mean obtained from single-line measurements from co-added spectra, there is good agreement between the \bz~measured here and the measurements reported by \citet{2012MNRAS.425.1278W} (their Table 6).

Measurements from individual lines are given in Table \ref{tab:line_bz}. Notably, there are few statistically significant differences in results from He lines, the C~{\sc iv} $\lambda$5801 line, or the O~{\sc iii} $\lambda$5592 line. This 
suggests that wind contamination in these lines is indeed minimal (or at least, at a similarly low level). Exceptions are He~{\sc i} $\lambda$4713, and He~{\sc ii} $\lambda$5412, which both yield lower amplitudes than other lines, likely due to wind contamination. The similarity of results across C, O, and most He lines furthermore confirms that Stokes $V$ is not affected by the presence of chemical spots, which can lead to large discrepancies in \bz~measurements obtained from different chemical species in cooler stars with Bp type chemical peculiarities \citep[e.g.][]{2015MNRAS.447.1418Y,2015MNRAS.449.3945S,2018MNRAS.475.5144S}. 
This is not surprising, as the strong winds of O-type stars are expected to render atomic diffusion inefficient in these stars by removing the outer layer where mass segregation occurs, inhibiting the formation of surface chemical spots (e.g. \citealt{1987ApJ...322..302M}). 

\begin{figure*}
\includegraphics[width=0.49\textwidth]{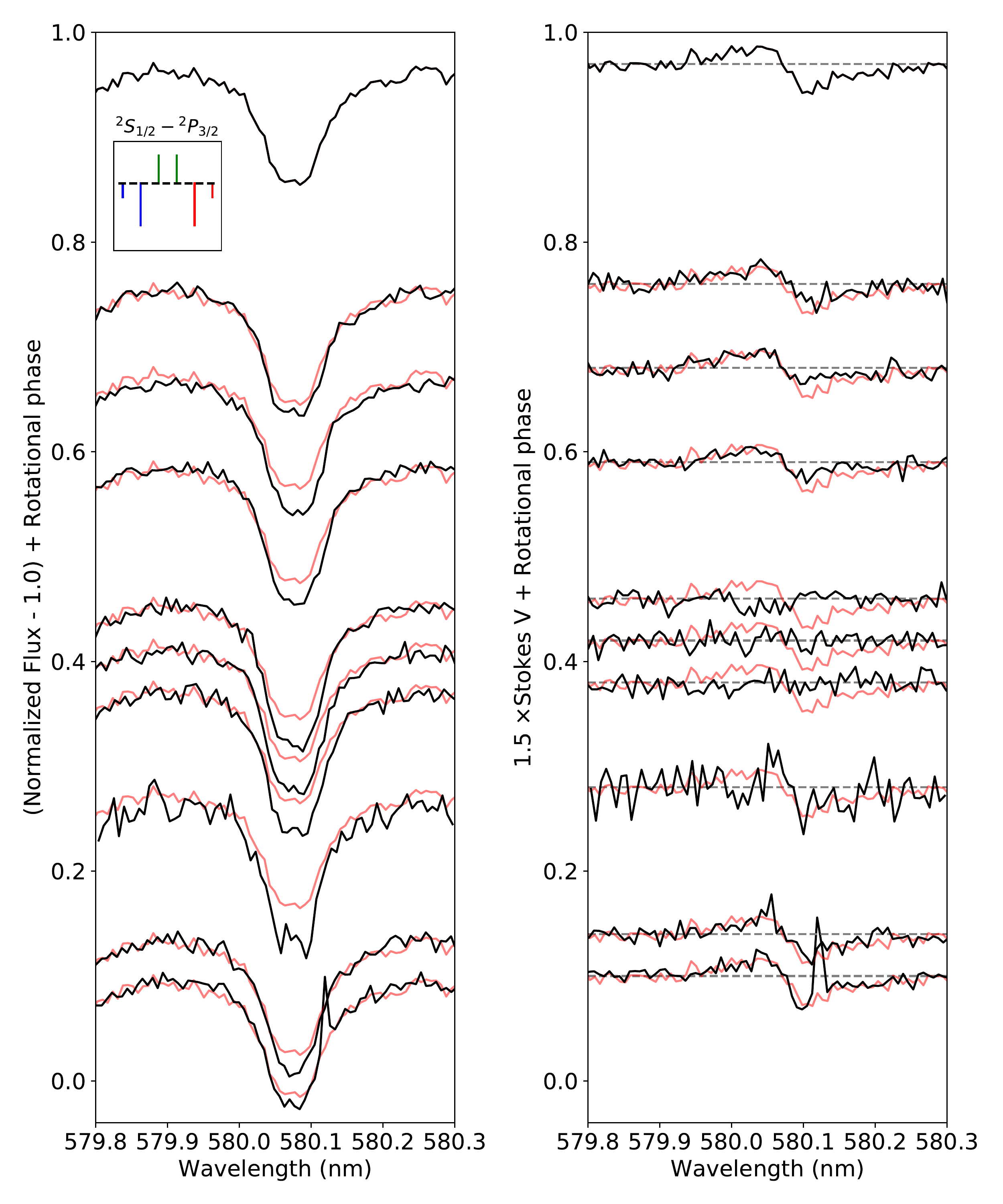}
\includegraphics[width=0.49\textwidth]{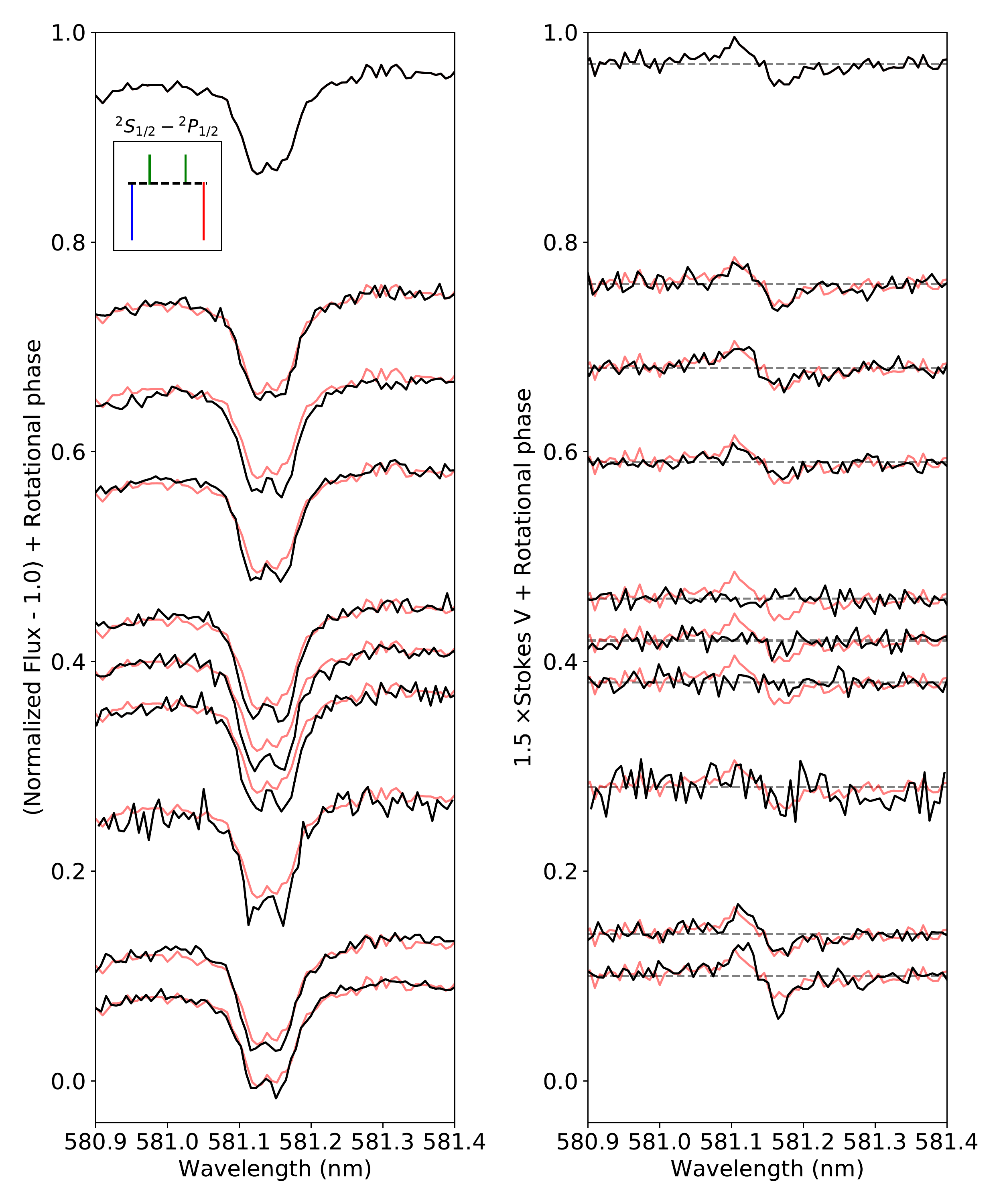}
\caption{
Spectropolarimetric dataset with vertical offset according to the rotational phase, for the C~\textsc{iv} $\lambda$5801 (left two panels) and 
C~\textsc{iv} $\lambda$5812 (right two panels) lines. The spectra were binned (2-pixel bins) for display purposes. The Zeeman patterns (on the same horizontal and vertical scales) are shown in the insets. The spectrum of the observations closest to phase 0 is overlaid with each subsequent spectrum, in pale red, to highlight the variation with phase. }
\label{CIV_stokes}
\end{figure*}

On the other hand, 
\bz~measurements made using the C {\sc iv} $\lambda$5812 line 
exhibit clear departures from the other lines. This is almost certainly a result of the high magnetic sensitivity of this line, which causes its shape to be broadened by the field and affects the measurements. Indeed, as pointed out by \citet{2012MNRAS.425.1278W}, Zeeman splitting (which scales with the Land\'{e} factor of each line) is visible in the C~\textsc{iv} $\lambda$5801 and $\lambda$5812 lines. Fig.~\ref{CIV_stokes} shows the profile variations over phase for both lines. The spectrum 
nearest to the high state (phase 0.97) is overlaid with each spectrum to highlight the phase variation. We notice that magnetic splitting is seen at all phases in the C~\textsc{iv} $\lambda$5812 line.

The fifth panel of Fig.\ \ref{bz_fwhm} shows the Full-Width at Half-Maximum (FWHM) of C~{\sc iv} $\lambda$5801 and $\lambda$5812 folded with 
with the same rotational period (157.99 d) used for the bottom panel, as a proxy 
of the magnetic modulus. A direct measurement of the modulus, by fitting synthetic profiles to the observed lines, is made challenging by the large degeneracies 
between the magnetic, thermal and turbulent broadening mechanisms, since only a few 
km~s$^{-1}$ of extra turbulent broadening, for instance, leads to large differences in the inferred modulus. Nevertheless, the FWHM of these C~{\sc iv} lines 
appears to exhibit rotational modulation: the reduced $\chi^2$ of a harmonic fit to the C~\textsc{iv} $\lambda$5812 line is 1.3, as compared to 2.9 for the null hypothesis of no variation about the mean value, and its FWHM peaks at around phase 0.5. While the null hypothesis cannot quite be formally ruled out (at a 99 per cent confidence threshold), this result still strongly suggests that rotational modulation is present, a result 
that could be better confirmed with higher S/N observations. By contrast, the less magnetically sensitive C~{\sc iv} $\lambda$5801 line\footnote{Though the difference between the effective Land\'{e} factor of this line (1.167) and that of the C~\textsc{iv} $\lambda$5812 line (1.333) is not that large, the Zeeman pattern of this line is also more complex (6 components) and less well resolved, probably leading to a more marginal variation of the magnetic broadening. Detailed modelling of the line profile would be required to include this line in the study of the magnetic field modulus.} shows 
an anti-correlated variation, likely due to the fact that its wings and width appear to change only minimally, while its depth varies, leading to a smaller FWHM when the line is deeper (e.g. around phase 0.5, as can be seen in Fig.~\ref{CIV_stokes}). Since there is no reason to expect that turbulent or thermal broadening varies horizontally across the stellar surface, the variation in the FWHM of the C~\textsc{iv} $\lambda$5812 line is most likely due to variation in the magnetic modulus.

Curiously, the FWHM of the C~{\sc iv} $\lambda$5812 line -- and therefore also the magnetic modulus, assuming that all variations in the FWHM result from variations in the modulus -- is at a maximum near phase 0.5, at which phase \bz~is consistent with 0. This behaviour is not 
expected for a dipolar magnetic field, as the magnetic modulus at the equator of a dipole should instead be at its minimum, but is commonly seen in Ap stars, for which the measured field modulus variations can be reproduced by adding a quadrupolar component to the global field topology (e.g., \citealt{2000A&A...359..213L}). Further indication that the field is probably not purely dipolar is given by the harmonic fit to the mean \bz~measurements (calculated from all lines except the discrepant C~{\sc iv} $\lambda$5812 line). A first-order fit, as would be appropriate to a dipolar field, yields a reduced $\chi^2$ of 10.8, as compared to 4.5 when \bz~is fit with 
three harmonics (Fig.\ \ref{bz_fwhm}, bottom).


\section{Discussion and conclusions}\label{sec:concl}

\subsection{Rotational period}

The first thing to rule out as a means of interpreting the variation of the UV spectrum is the possibility that the rotational period, calculated using equivalent width measurements of a few lines and line ratios by \citet{2012MNRAS.425.1278W}, is wrong. While they already provide compelling arguments as to why other periods are not suitable to reproduce the variations of these lines, we 
further 
investigate whether a different rotational period would help make sense of the UV observations presented above.

If the current rotational period is wrong, while spectra with similar line morphologies might not have been obtained at exactly the same rotational phase, within a dipolar field framework, one would imagine that they were obtained at relatively close phases (within a difference of up to $\sim$0.2 cycles). In particular, the spectra obtained at phases 0.27 and 0.87 (with respect to the published ephemerides) are nearly identical in character, yet they were obtained only $\sim$60 days apart. This could suggest either that the period is significantly longer (or shorter) than the published value.

First, 
assuming for instance that the equivalent widths of the aforementioned lines undergo a double-wave variation over a single rotational cycle, one might surmise that the ``real" period is in fact 315.98 d, or twice the previously inferred value. 
Such a period would encompass all of our new UV observations, 
and while the 
UV line profiles phased with this longer period appear to vary in a more coherent manner, they do so in a single-wave pattern, which is inconsistent with the global picture derived from other observational diagnostics, 
which would then show a double-wave variation 
in this assumed scenario.

On the other hand, if we consider the possibility of the period being shorter (perhaps by a factor of 2 or 3, given the UV observations), we 
encounter further complications. Not only are significantly shorter periods ruled out by the equivalent width analysis performed by \citet{2012MNRAS.425.1278W}
, down to 60~d, but they are also difficult to reconcile with the profiles of photospheric lines in the optical spectra of NGC 1624-2. \citet{2012MNRAS.425.1278W} place an upper limit on $v \sin i$ of 15 km\,s$^{-1}$, based on the modelling of the C~\textsc{iv} $\mathrm{\lambda}$5801 line, assuming macroturbulent 
broadening, but without magnetic broadening. Conversely, when they model this line (which is quite magnetically sensitive due to its high Land\'{e} factor) using Zeeman broadening, they find their best fit to be consistent with all of the broadening being accounted for by the Zeeman effect (as discussed above), or essentially no rotational broadening. Based on their 
estimated values for the stellar radius ($R_* = 10 R_\odot$) and their measured rotation period, they find an upper limit on the equatorial rotational velocity of $\sim$3.2 km\,s$^{-1}$. Increasing that value by a factor of 2 or more requires us to adopt a fairly low value of the inclination angle $i$, which does not appear to be consistent with the large variations seen in the UV. The magnetic measurements presented 
in Figs.~\ref{bz_fwhm} and \ref{CIV_stokes} also phase well with the published period, and rule out 
a period that is significantly different from that value, or possibly its double (in the case of a double-wave variation of the longitudinal field).

While we cannot, within the scope of this article, constrain a firm lower limit on the inclination based on UV variations alone, 
further modelling of the surface magnetic field, including 
modelling of the photospheric line profiles, is in progress and should conclusively resolve this issue. In the meantime, however, we infer that the previously derived rotational period is 
probably correct, or sufficiently close 
to the 
\textit{real} value that it does not affect the interpretation that similar line morphologies are 
observed at three separate rotational phases
.

\subsection{Magnetospheric structure and surface magnetic geometry}\label{sec:mag}

The most obvious hypothesis to attempt to explain the puzzling behaviour of the UV resonance lines of NGC 1624-2 involves a complex magnetospheric structure, with significant departures from what would be expected in the case of a purely dipolar surface magnetic field interacting with the stellar wind. Our preliminary magnetic analysis shows that the surface magnetic field of NGC 1624-2 is likely better modelled using a more complex geometry
, though 
the determination of the detailed magnetic geometry is outside the scope of this paper. While this explanation appears satisfying from the point of view of these observations, it also appears at odds with the behaviour of optical lines \citep{2012MNRAS.425.1278W}, in particular He~\textsc{ii} $\lambda$4686 (their Fig.~5), which despite certain gaps in phase coverage, seems to behave in a fairly monotonic way between high state (phase 0.0) and low state (phase 0.5), and vice versa. This behaviour is more in line with what is observed in the optical lines of other magnetic O stars, whose fields are inferred to be dipolar, such as HD~191612 \citep{2013MNRAS.431.2253M}.

One 
possibility is that the explanation could reside not necessarily in the complexity of the field responsible for the formation of the magnetosphere, but instead arise as a consequence of dynamic flows within the magnetosphere, predicted in MHD simulations to occur (e.g. \citealt{2002ApJ...576..413U}). It could be that some observational diagnostics -- especially those formed further away from the stellar surface, such as the 
UV resonance lines -- are more sensitive to these flows than others, which are dominated by averaging effects. However, while this could explain some discrepancies at low velocities, it is difficult to produce high-velocity blueshifted absorption 
(as observed in the C~\textsc{iv} $\lambda\lambda$1548/50 doublet) 
using these dynamic flows. In fact, investigations of different implementations of the ADM model using material either flowing up or down closed field loops (mimicking in a way the aforementioned flows) do not lead to any variations in the line profile past a velocity of about 70 per~cent of the inferred terminal velocity \citep{2018A&A...616A.140H}. The increased depth of the C~\textsc{iv} doublet at high velocity at rotational phases 0.274, 0.621 and 0.986 suggest that we are looking down 
open field lines, something that can only occur, as far as we can tell, if the field is topologically complex (given that this line is understood to form throughout most of the wind, therefore probing its full structure).

While the detailed 
geometry of the surface magnetic field cannot be 
inferred from the effect that it has on the magnetospheric structure and hence on the variation of UV resonance line profiles, better spectropolarimetric phase coverage and surface mapping techniques such as Zeeman Doppler Imaging (ZDI; \citealt{1989A&A...225..456S, 2002A&A...381..736P}) have the potential of shedding light on NGC 1624-2's potentially complex magnetic field. This topic will be addressed in more detail in a forthcoming publication.

\subsection{Magnetic field origin and evolution}

A photometric study of the NGC 1624 cluster has yielded an upper limit on the main-sequence turnoff age of 4 Myr \citep{2015AJ....149..127L}. 
A better characterization of this cluster (and in particular of its massive star population by means of quantitative spectroscopy) would prove useful to get a more precise age and better constrain the evolutionary status of NGC 1624-2.

This question is of particular relevance in the context of the origin of magnetic fields 
in massive stars and their evolution. It has been suggested that their fields might decay over time (e.g. \citealt{2007A&A...470..685L, 2008A&A...481..465L, 2016A&A...592A..84F}), and that the decay rate increases with higher mass, from near-flux conservation in A stars \citep{2019MNRAS.483.3127S} to stronger flux decay in OB stars \citep{2019MNRAS.490..274S}. Assuming Ohmic dissipation to be the mechanism by which these fields decay, higher order multipolar components are expected to decay fastest (e.g. \citealt{2013SAAS...39.....C}, and references therein; there is also tentative evidence that complex fields decay faster than simple dipolar fields in OB stars, \citealt{2019MNRAS.490..274S}). Therefore, a strong, complex field might be consistent with the Ohmic dissipation scenario (and indicative of a young magnetic star, if its field is fossil in origin), and a more detailed characterization of the field topology of magnetic high-mass stars might prove crucial to reveal the phenomena influencing the evolution of magnetism in these stars.

Alternatively, stellar mergers have 
been proposed as a viable channel for the formation of magnetic fields in massive stars \citep{2009MNRAS.400L..71F,2016MNRAS.457.2355S,2019Natur.574..211S, 2020MNRAS.495.2796S}. While merger products are expected to potentially exhibit large surface enhancements of hydrogen-burning products such as helium and nitrogen \citep{2013MNRAS.434.3497G}, NGC 1624-2 is believed to show at most only a moderate nitrogen surface enrichment \citep{2012MNRAS.425.1278W}, though the precise abundance is difficult to determine due partly to the uncertainties on the strength of the magnetic field and the effective temperature. A more thorough evaluation of the merger hypothesis in the specific case of NGC 1624-2 can only be conducted once its environment is better characterised, including especially abundance analyses of other massive stars in the cluster.

\subsection{Conclusions}

The line profile variations observed in \textit{HST}/\textit{COS} spectra of NGC 1624-2 over six separate rotational phases do not correspond to the expected phenomenology -- as established by observations of other magnetic O-type stars, or modelled using MHD simulations or the ADM prescription -- which accounts 
for a magnetosphere formed by the interaction of a strong stellar wind and a global dipolar field
. Instead, the observed line profiles appear to exhibit one of two morphologies, previously understood -- within the aforementioned scenario -- to correspond to either a magnetic pole-on or magnetic equator-on view of the magnetosphere. Furthermore, a magnetic analysis using spectropolarimetric observations shows that the diagnostics of NGC 1624-2's surface magnetic field cannot be well reproduced using a purely dipolar geometry, assuming that the FWHM of the C~\textsc{iv} $\lambda$5812 line acts as an appropriate proxy for the field modulus. Having exhausted more mundane explanations such as an error in the determination of the rotation period (such that our rotational phase determinations are sufficiently wrong to change the overall phenomenological picture) and short-term stochastic variations, we conclude that the magnetosphere of NGC 1624-2 is likely more structured than previously assumed, 
and that its surface magnetic field 
might depart 
from a dipolar geometry. While challenges remain in trying to reconcile such a scenario with other diagnostics (e.g. the smooth variations with phase of the equivalent widths of optical emission lines), only further modelling accounting for the detailed geometry of the field and the resulting distribution of plasma within the magnetosphere will allow us to self-consistently reproduce diagnostics ranging across multiple wavelength bands. Nevertheless, the 
perspective of the strongest known magnetic field on an O-star also being 
topologically complex provides us with crucial information regarding the formation and evolution of such fields on massive stars and further emphasizes the singular importance of this archetypal object.

\section*{Acknowledgements}

Some of the data presented in this paper were obtained from the Mikulski Archive for Space Telescopes (MAST). This work has made use of the VALD database, operated at Uppsala University, the Institute of Astronomy RAS in Moscow, and the University of Vienna. 
ADU and VP acknowledge support for this work through program HST-GO-15066.001-A that was provided by NASA through a grant from the Space Telescope Science Institute, which is operated by the Association of Universities for Research in Astronomy, Inc., under NASA contract NAS 5-26555.

ADU and GAW gratefully 
acknowledge the support of the Natural Sciences and Engineering Research Council of Canada (NSERC). VP acknowledges support from the University of Delaware Research Foundation. This material is
based upon work supported by the National Science Foundation under Grant No. 1747658. MES acknowledges support from the Annie Jump Cannon Fellowship, supported by the University of Delaware and endowed by the Mount Cuba Astronomical Observatory. CE acknowledges graduate assistant salary support
from the Bartol Research Institute in the Department of
Physics, University of Delaware, as well as support from
program HST-GO-13629.002-A that was provided by NASA
through a grant from the Space Telescope Science Institute.

Finally, the authors also thank D.~H. Cohen, as well as the anonymous referee of this manuscript, for useful discussions and feedback which contributed to improve this paper.

\section*{Data Availability}

The 
ultraviolet spectroscopic data underlying this article are available in the Mikulski Archive for Space Telescopes (MAST) at \url{https://archive.stsci.edu/}, and are uniquely identified with the observation identifiers (ObsID) listed in Table~\ref{tab:obs}. The optical spectropolarimetric data underlying this article are available in the Canadian Astronomy Data Center (CADC) archive at \url{https://www.cadc-ccda.hia-iha.nrc-cnrc.gc.ca/en/} and the PolarBase database \citep{1997MNRAS.291..658D,2014PASP..126..469P} available at \url{http://polarbase.irap.omp.eu/}, and are uniquely identified with the ObsIDs listed in Table~\ref{tab:Obs_mag}.




\bibliographystyle{mnras}
\bibliography{database} 




\appendix

\section{Journal of observations}

In this section, we list the \textit{HST}
, CFHT and TBL observations used in the context of this study, as well as the longitudinal magnetic field measurements performed using the latter two.

\begin{table*}
\caption[c]{Journal of COS Observations}\label{tab:obs}
\begin{tabular}{@{}llccccccc}
\hline
ObsID     & Program ID & Grating & {${\lambda_c}^1$} & S/N$^{2}$ & UT (Start)         & Exp. Time & MJD(mid)   & $\phi\,^3$ \\
          &    &     &    (\AA)  &        &                    &  (s)      &            &            \\
\hline                 
lcl601010 & 13734 & G130M  &  1291 & 10.6           & 2015-02-17T06:16:42 &  ~848     & 57070.2685 & 0.986      \\
lcl601020 & 13734 & G130M  &  1327 & 10.1           & 2015-02-17T06:39:53 &  ~844     & 57070.2846 & 0.986      \\
lcl601030 & 13734 & G160M  &  1577 & 16.1           & 2015-02-17T07:40:05 &  1084     & 57070.3279 & 0.986      \\       
lcl601040 & 13734 & G160M  &  1623 & 16.0           & 2015-02-17T08:07:48 &  1080     & 57070.3472 & 0.986      \\   
lcl602010 & 13734 & G130M  &  1291 & 10.2           & 2015-10-06T16:20:53 &  ~848     & 57301.6881 & 0.451      \\  
lcl602020 & 13734 & G130M  &  1327 & 9.9           & 2015-10-06T16:44:04 &  ~844     & 57301.7042 & 0.451      \\ 
lcl602030 & 13734 & G160M  &  1577 & 15.6           & 2015-10-06T17:07:58 &  1084     & 57301.7309 & 0.451      \\
lcl602040 & 13734 & G160M  &  1623 & 15.6           & 2015-10-06T18:00:30 &  1080     & 57301.7588 & 0.452      \\
ldmu04010 & 15066 & G130M & 1222 & 10.5 & 2018-02-08T09:59:05 & 940 & 58157.4233 & 0.867 \\
ldmu04020 & 15066 & G130M & 1291 & 12.0 & 2018-02-08T10:23:13 & 940 & 58157.4390 & 0.868 \\
ldmu04030 & 15066 & G160M & 1577 & 15.7 & 2018-02-08T11:25:11 & 1080 & 58157.4842 & 0.868 \\
ldmu04040 & 15066 & G160M & 1623 & 15.6 & 2018-02-08T11:52:50 & 1080 & 58157.5033 & 0.868 \\
ldmu01010 & 15066 & G130M & 1222 & 10.4 & 2018-04-13T13:44:46 & 940 & 58221.5927 & 0.274 \\
ldmu01020 & 15066 & G130M & 1291 & 11.9 & 2018-04-13T14:45:31 & 940 & 58221.6212 & 0.274 \\
ldmu01030 & 15066 & G160M & 1577 & 15.4 & 2018-04-13T15:20:10 & 1080 & 58221.6605 & 0.274 \\
ldmu01040 & 15066 & G160M & 1623 & 15.6 & 2018-04-13T16:25:40 & 1080 & 58221.6986 & 0.274 \\
ldmu02010 & 15066 & G130M & 1222 & 10.4 & 2018-10-01T21:47:01 & 940 & 58392.9149 & 0.358 \\
ldmu02020 & 15066 & G130M & 1291 & 12.0 & 2018-10-01T23:00:25 & 940 & 58392.9648 & 0.358 \\
ldmu02030 & 15066 & G160M & 1577 & 15.6 & 2018-10-01T23:22:25 & 1080 & 58393.0032 & 0.359 \\
ldmu02040 & 15066 & G160M & 1623 & 15.8 & 2018-10-02T00:50:19 & 1080 & 58393.0432 & 0.359 \\
ldmu03010 & 15066 & G130M & 1222 & 10.6 & 2018-11-12T11:23:02 & 940 & 58434.4816 & 0.621 \\
ldmu03020 & 15066 & G130M & 1291 & 12.0 & 2018-11-12T12:30:57 & 940 & 58434.5277 & 0.622 \\
ldmu03030 & 15066 & G160M & 1577 & 15.9 & 2018-11-12T12:52:57 & 1080 & 58434.5452 & 0.622 \\
ldmu03040 & 15066 & G160M & 1623 & 15.8 & 2018-11-12T14:06:15 & 1080 & 58434.5959 & 0.622 \\
\hline
\multicolumn{7}{l}{$^1$ Central wavelength of the grating setting.}                                    \\
\multicolumn{9}{l}{$^2$ Computed per 9-pixel resolution element between 1350--1355~{\AA} (G130M) or 1490--1495~{\AA} (G160M). }     \\
\multicolumn{9}{l}{$^3$ Phase of MJD(mid) according to the ephemeris of Wade et al. (2012). }\\

\end{tabular}
\end{table*}

\begin{table*}
\begin{center}
\caption[ List of observations of NGC 1624-2]{Observations of NGC 1624-2 from ESPaDOnS at CFHT and Narval at TBL (in italic).  In this table HJD is the Heliocentric Julian Date, and JD0 is the arbitrary Julian Date chosen as the starting point to record the rotations of NGC~1624-2.  The rotation 
numbers and phases are based on JD0 and NGC 1624-2's well defined rotation period of 157.99 days. }\label{tab:Obs_mag}
\begin{tabular}{l l l c l l c c}%
\hline
Date & ObsID & HJD & Rotations & Phase & Mean Phase & Mean  \bz & Mean \nz \\
     &       &-2,450,000 & since JD0 & & & (kG) & (kG) \\
\hline
2012-02-01 &      1522987p & 5958.7154  &   -1  &  0.948 $\pm$ 0.06   &         0.97 $\pm$ 0.06 & $5.10 \pm 0.22$ & $0.13 \pm 0.22$ \\
2012-02-02 &      1523106p & 5959.7155  &   -1  &  0.954 $\pm$ 0.06   &   & &       \\
2012-02-03 &      1523413p & 5960.7132  &   -1  &  0.960 $\pm$ 0.06   &   & &       \\
2012-02-04 &      1523605p & 5961.7128  &   -1  &  0.966 $\pm$ 0.06   &     & &     \\
2012-02-09 &      1524145p & 5966.7204  &   -1  &  0.998 $\pm$ 0.06   &  & &         \\
\hline
\textit{2012-03-24} &      \textit{119496o}  & \textit{6011.3316}  &   \textit{0}   &  \textit{0.281 $\pm$ 0.06}   & \textit{0.28 $\pm$ 0.06} & \textit{3.1 $\pm$ 0.6} &  \textit{0.1 $\pm$ 0.6}   \\
\hline
2012-09-27 &      1573465p & 6197.9445  &   1  &  0.462 $\pm$ 0.07   &         0.46 $\pm$ 0.07 & $-0.49 \pm 0.21$ & $-0.38 \pm 0.21$ \\
2012-09-27 &      1573469p & 6198.0090  &   1  &  0.462 $\pm$ 0.07   & & &         \\
\hline
2013-08-27 &      1650338p & 6532.1192  &   3  &  0.577 $\pm$ 0.09   &         0.59 $\pm$ 0.09 & $2.67 \pm 0.18$ & $0.23 \pm 0.19$ \\
2013-08-29 &      1650656p & 6534.0561  &   3  &  0.589 $\pm$ 0.09   &  & &       \\
2013-08-29 &      1650660p & 6534.1180  &   3  &  0.590 $\pm$ 0.09   &  & &       \\
\hline
2013-09-13 &      1653352p & 6549.0333  &   3  &  0.684 $\pm$ 0.09   &         0.68 $\pm$ 0.09 & $3.75 \pm 0.23$ & $-0.00 \pm 0.23$ \\
2013-09-13 &      1653356p & 6549.0961  &   3  &  0.684 $\pm$ 0.09   &  & &        \\
\hline
2013-09-25 &      1656332p & 6561.0236  &   3  &  0.760 $\pm$ 0.09   &         0.76 $\pm$ 0.09 & $4.46 \pm 0.28$ & $0.53 \pm 0.28$ \\
2013-09-25 &      1656336p & 6561.0859  &   3  &  0.760 $\pm$ 0.09   &  & &        \\
\hline
2013-11-17 &      1669969p & 6613.9090  &   4  &  0.095 $\pm$ 0.09   &         0.10 $\pm$ 0.09 & $4.24 \pm 0.24$ & $-0.14 \pm 0.24$ \\
2013-11-17 &      1669981p & 6614.0410  &   4  &  0.095 $\pm$ 0.09   &  & &        \\
\hline
2013-11-24 &      1671235p & 6621.0356  &   4  &  0.140 $\pm$ 0.09   &         0.14 $\pm$ 0.09 & $3.97 \pm 0.27$ & $0.50 \pm 0.27$ \\
2013-11-24 &      1671239p & 6621.0978  &   4  &  0.140 $\pm$ 0.09   &    & &      \\
\hline
2014-01-08 &      1682621p & 6665.8216  &   4  &  0.423 $\pm$ 0.09   &         0.42 $\pm$ 0.09 & $0.37 \pm 0.28$ & $-0.32 \pm 0.28$ \\
2014-01-08 &      1682625p & 6665.8835  &   4  &  0.424 $\pm$ 0.09   &  & &        \\
\hline
2015-09-24 &      1834775p & 7290.0362  &   8  &  0.374 $\pm$ 0.11   &         0.38 $\pm$ 0.11 & $-0.43 \pm 0.27$ & $-0.25 \pm 0.27$ \\
2015-09-25 &      1834930p & 7291.0543  &   8  &  0.381 $\pm$ 0.11   &      & &    \\
\hline
\end{tabular}
\end{center}
\end{table*}

\begin{table*}
\begin{center}
\caption[]{\bz~measurements from individual lines, in kG. The second row under each line of the table heading gives the Land\'e factor of each line, as found in the output of the VALD database, or computed from LS coupling. In the case of multiplets, the highest value among the components of the line complex is shown.}
\label{tab:line_bz}
\begin{tabular}{l c c c c c c c c c c}
\hline\hline
HJD - & He{\sc I} 4471 & He{\sc I} 4713 & He{\sc I} 4922 & He{\sc I} 5016 & He{\sc I} 7281 & He{\sc II} 5412 & C{\sc IV} 5801 & C{\sc IV} 5812 & O{\sc III} 5592 \\
2,450,000 & 1.333 & 1.750 & 1.000 & 1.200 & 1.000 & 1.200 & 1.167 & 1.333 & 1.000 \\
\hline
5961.5156  & $ 6.7 \pm  1.7$  & $ 3.3 \pm  0.5$  & $ 6.0 \pm  0.6$  & $ 3.7 \pm  0.4$  & $ 4.8 \pm  0.4$  & $ 4.0 \pm  0.5$  & $ 6.2 \pm  0.5$  & $ 4.0 \pm  0.6$  & $ 6.2 \pm  0.9$  \\
6011.3316  & $ 1.0 \pm  3.1$  & $ 2.9 \pm  1.9$  & $ 2.9 \pm  1.9$  & $ 2.1 \pm  1.1$  & $ 2.8 \pm  1.1$  & $ 4.8 \pm  1.6$  & $ 3.0 \pm  1.3$  & $ 3.0 \pm  1.8$  & $ 2.3 \pm  2.9$  \\
6197.9766  & $-0.1 \pm  1.3$  & $-0.0 \pm  0.5$  & $-0.6 \pm  0.7$  & $-0.2 \pm  0.3$  & $-1.0 \pm  0.4$  & $-0.7 \pm  0.7$  & $-0.7 \pm  0.6$  & $ 0.0 \pm  0.5$  & $-0.3 \pm  1.0$  \\
6533.4312  & $ 1.1 \pm  1.3$  & $ 1.5 \pm  0.6$  & $ 1.6 \pm  0.7$  & $ 2.1 \pm  0.4$  & $ 3.4 \pm  0.4$  & $ 2.0 \pm  0.5$  & $ 2.9 \pm  0.4$  & $ 2.0 \pm  0.4$  & $ 2.5 \pm  0.9$  \\
6549.0645  & $ 4.4 \pm  1.6$  & $ 1.6 \pm  0.7$  & $ 3.0 \pm  0.9$  & $ 2.9 \pm  0.3$  & $ 4.1 \pm  0.5$  & $ 3.6 \pm  0.6$  & $ 4.7 \pm  0.5$  & $ 3.8 \pm  0.6$  & $ 3.0 \pm  0.8$  \\
6561.0547  & $ 7.4 \pm  2.2$  & $ 2.3 \pm  0.7$  & $ 4.5 \pm  1.0$  & $ 3.6 \pm  0.5$  & $ 5.3 \pm  0.5$  & $ 3.7 \pm  0.8$  & $ 4.6 \pm  0.7$  & $ 2.9 \pm  0.6$  & $ 4.1 \pm  1.0$  \\
6613.9751  & $ 0.2 \pm  1.8$  & $ 2.6 \pm  0.6$  & $ 3.4 \pm  0.7$  & $ 3.2 \pm  0.3$  & $ 5.3 \pm  0.4$  & $ 4.5 \pm  0.8$  & $ 4.1 \pm  0.8$  & $ 4.3 \pm  0.6$  & $ 4.6 \pm  1.4$  \\
6621.0664  & $ 1.2 \pm  1.4$  & $ 2.0 \pm  0.8$  & $ 5.2 \pm  0.8$  & $ 3.1 \pm  0.4$  & $ 5.5 \pm  0.6$  & $ 3.8 \pm  0.7$  & $ 3.0 \pm  0.8$  & $ 3.3 \pm  0.5$  & $ 5.4 \pm  1.0$  \\
6665.8525  & $ 1.1 \pm  1.8$  & $ 0.2 \pm  0.8$  & $-0.2 \pm  0.9$  & $ 0.8 \pm  0.6$  & $ 0.9 \pm  0.5$  & $-0.3 \pm  0.7$  & $-0.4 \pm  0.7$  & $ 0.9 \pm  0.6$  & $-1.8 \pm  1.2$  \\
7290.5449  & $ 1.6 \pm  2.0$  & $-0.9 \pm  0.6$  & $-1.1 \pm  1.0$  & $-0.4 \pm  0.4$  & $ 0.2 \pm  0.5$  & $ 0.9 \pm  0.9$  & $-1.9 \pm  0.7$  & $ 0.4 \pm  0.6$  & $ 0.5 \pm  1.1$  \\
\hline
\end{tabular}
\end{center}
\end{table*}


\bsp	
\label{lastpage}
\end{document}